\documentclass[conference]{IEEEtran}
\IEEEoverridecommandlockouts
\usepackage{cite}
\usepackage{amsmath,amssymb,amsfonts}
\usepackage{algorithmic}
\usepackage{graphicx}
\usepackage{textcomp}
\usepackage{xcolor}
\def\BibTeX{{\rm B\kern-.05em{\sc i\kern-.025em b}\kern-.08em
    T\kern-.1667em\lower.7ex\hbox{E}\kern-.125emX}}

\usepackage{dirtytalk}
\usepackage{algorithm2e}
\usepackage{amsmath}
\usepackage{amssymb}
\usepackage{bm}
\usepackage{pifont}% http://ctan.org/pkg/pifont
\usepackage[colorinlistoftodos]{todonotes}
\newcommand{\cmark}{\ding{51}}%
\newcommand{\xmark}{\ding{55}}%
\usepackage{dsfont}
\usepackage{booktabs}
\usepackage{cite}
\usepackage{hyperref}

\begin{document}

\title{Stock Embeddings: Learning Distributed Representations for Financial Assets
%-- UPDATE ANONYMOUS GITHUB 
\thanks{This publication has emanated from research conducted with the financial support of Science Foundation Ireland under Grant number 18/CRT/6183. For the purpose of Open Access, the author has applied a CC BY public copyright licence to any Author Accepted Manuscript version arising from this submission.}
}

% \author{\IEEEauthorblockN{Anonymous Authors}}

%-- UPDATE ANONYMOUS GITHUB AND FUNDING

\author{\IEEEauthorblockN{Rian Dolphin}
\IEEEauthorblockA{\textit{School of Computer Science} \\
\textit{University College Dublin}\\
Dublin, Ireland \\
rian.dolphin@ucdconnect.ie}
\and
\IEEEauthorblockN{Barry Smyth}
\IEEEauthorblockA{\textit{School of Computer Science} \\
\textit{University College Dublin}\\
Dublin, Ireland \\
barry.smyth@ucd.ie}
\and
\IEEEauthorblockN{Ruihai Dong}
\IEEEauthorblockA{\textit{School of Computer Science} \\
\textit{University College Dublin}\\
Dublin, Ireland \\
ruihai.dong@ucd.ie}
}

\maketitle

\begin{abstract}
Identifying meaningful relationships between the price movements of financial assets is a challenging but important problem in a variety of financial applications. However with recent research, particularly those using machine learning and deep learning techniques, focused mostly on price forecasting, the literature investigating the modelling of asset correlations has lagged somewhat. To address this, inspired by recent successes in natural language processing, we propose a neural model for training stock embeddings, which harnesses the dynamics of historical returns data in order to learn the nuanced relationships that exist between financial assets. We describe our approach in detail and discuss a number of ways that it can be used in the financial domain. Furthermore, we present the evaluation results to demonstrate the utility of this approach, compared to several important benchmarks, in two real-world financial analytics tasks.
\end{abstract}
%on harnessing the dynamics of historical price for 
%Identifying meaningful relationships between the price movements of financial assets is a challenging but important problem in a variety of financial applications. However with recent research, particularly those using machine learning techniques, focused mostly on price forecasting, literature investigating the modelling of asset correlations has lagged somewhat. To address this, inspired by recent successes in natural language processing, we propose a novel model for training stock embeddings using historical returns data in order to capture the nuanced relationships that exist between financial assets. We describe our approach in detail and discuss a number of ways that it can be used in the financial domain. Furthermore, we present the results of an initial evaluation to demonstrate the utility of this approach, compared to several benchmarks, in two real-world financial analytics tasks.


\section{Introduction}

The stock market provides a challenging but appealing target for predictive analysis~\cite{bachelier1900theorie,fama1970efficient,fama1965behavior}, and has attracted the attention of machine learning (ML) research for some time. While state-of-the art ML techniques, such as modern deep learning-based methods, have demonstrated excellent performance on financial forecasting tasks \cite{sezer2020financial}, less attention has been paid to a range of other important tasks. One such area is portfolio optimisation, diversification and hedging. This relies heavily on being able to understand the relationships and similarities between financial assets; to identify groups of assets that will perform well collectively, for example by delivering a stable and predictable increase in value over time with reduced risk of capital loss.  It is this problem of measuring relationships and similarities between financial assets that we seek to address in this paper.

Allocating capital within an investment portfolio is a two-step process~\cite{markowitz1952portfolio}, \emph{forecasting} and \emph{allocation}: (1) an investor must form some beliefs about the future return and risk potential of each asset \emph{individually}; and (2) given their beliefs, the investor must then decide how to allocate capital between these assets taking into account the relationships and correlations that may exist between pairs of assets and their returns. Most relevant recent research has focused on the forecasting step, usually on an individual asset basis, by proposing models to predict asset returns and volatility~\cite{sezer2020financial}, with less attention paid to optimising the allocation of assets (\emph{portfolio design}); indeed modern portfolio theory largely remains rooted in \cite{markowitz1952portfolio}, which saw Harry Markowitz receive the Nobel Prize in Economics in 1990. 

In this paper, inspired by language models that have proven to be very successful in the NLP domain \cite{mikolov2013efficient, pennington2014glove,firth1957synopsis}, we propose a method for learning stock embeddings which are capable of capturing the underlying relationships between stocks by studying distributional similarities in their historical returns. In particular, we propose a novel algorithm and training methodology for the learning of distributed representations of financial assets, based purely on historical returns data, which can be used to identify financial assets that are likely to behave similarly and differently, both important elements of portfolio design. After summarising related work in the next section, we go on to describe this approach and then present a number of example case-studies to showcase how the learned representations might be useful in financial applications. Finally, before concluding, we present the results of an initial evaluation to demonstrate the effectiveness of the learned representations.

\section{Related Work}

The rapid advancement of ML in recent years has fueled interest in applying modern techniques within the financial domain, with applications ranging from tackling financial cybercrime~\cite{nicholls2021financial} to stock market forecasting 
~\cite{ding2015deep}. Within the subset of applications focusing on the stock market, the literature overwhelmingly focuses on return or volatility forecasting for individual assets~\cite{yang2020,ding2015deep}. Although these forecasting models have achieved improvements, there are many other problems within the stock market which have not received the same level of attention. One such area is the modelling of asset correlations and relationships.

Modelling relationships and correlations between asset returns is vital for a number of financial tasks such as portfolio optimisation, hedging and sector classification~\cite{phillips2016industry}. Markowitz used covariance of returns to inform a risk reduction strategy in his pioneering paper on modern portfolio theory~\cite{markowitz1952portfolio} and, as a result, correlation has become the conventional measure of similarity between the returns time series of two financial assets. 
A number of other techniques for measuring asset similarity have been proposed in recent years including a geometric measure of similarity~\cite{chun2020geometric} and an adjusted correlation-based approach~\cite{dolphin2021measuring}. As previously mentioned, ML based solutions to this task are lacking, with many relying on textual data as discussed below.

% However, his use of covariances is based on the underlying assumption that stock returns follow a normal distribution, something which has been proven to be false~\cite{hagerman1978more}. 
% Other researchers have since proposed alternative solutions to the portfolio optimisation problem, one example being Hierarchical Risk Parity~\cite{lopez2016HRP}, which uses machine learning and graph theory to replace the traditional covariance structure with a tree structure.

Though computing pairwise correlation or covariance estimates between assets is an actionable solution, we hypothesise that the idea of representing financial assets using dense vector embeddings has the potential to outperform traditional correlation based approaches. The learning of distributed representations has been extensively studied in recent years with arguably the most well-known application being word embeddings in language modeling~\cite{wang2020stock2vec}. Motivated by algorithms such as Word2Vec~\cite{mikolov2013efficient}, neural embedding methods have become more prevalent across multiple domains such as recommendation systems~\cite{barkan2016item2vec} and medicine~\cite{choi2016medical}.

The application of embeddings in the financial domain have, for the most part, been limited to the NLP idea of word embeddings. For example, \cite{ding2015deep, ding2016knowledge, cheng2020knowledge} tackle stock returns forecasting using event embeddings obtained from financial news. \cite{ito2020embedding} train company embeddings by applying BERT to the textual data from annual reports and \cite{hirano2018selection} use word embeddings to select similar stocks. Despite applying the idea of embeddings within the financial domain, the aforementioned literature still relies on the aggregation of pretrained word embeddings from language models, rather than on using a novel technique to learn embeddings from non-textual financial data such as historical returns.

In contrast, the method proposed in this paper does not draw on pretrained word embeddings or any form of textual data. %Instead we propose a novel methodology, inspired by language models from the NLP domain, but which we adapt to only use historical returns data, to learn stock embeddings.
Instead, we propose a novel methodology for generating stock embeddings using only historical returns data.

% \cite{tagarev2019comparison} - This paper compares a number of language models for the purpose of sector classification using 10-k data.

% \cite{li2020enhancing} - Enhancing Stock Trend Prediction Models by Mining Relational Graphs of Stock Prices

%==================================================
%==================================================
%==================================================
\section{Architecture \& Approach}\label{sec:Methodology}

This paper proposes a method for training distributed representations of stocks using a probabilistic neural framework. Inspired by the distributional semantics area of natural language processing, we propose a model architecture and training procedure that uses the idea of \emph{context stocks} to learn embeddings of target stocks. In linguistics, the distributional hypothesis, which underpins a number of popular language models~\cite{mikolov2013efficient}, captures the idea that \say{a word is characterized by the company it keeps}~\cite{firth1957synopsis}, i.e., words that occur in the same contexts tend to have similar meanings. 

In the financial domain, research has shown a similar hypothesis to be true: companies with similar characteristics, such as operating in the same business sectors, tend to exhibit similar stock price fluctuations~\cite{gopikrishnan2000sector_price}. As a result, by engineering the selection of context stocks to reflect this hypothesis, and adding noise reduction strategies, our proposed framework generates embeddings that can capture relationships between financial assets purely based on historical pricing data.

%We propose stocks that disproportionately exhibit similar returns at the same time are similar and this is the foundation for selection of \emph{context stocks}. This is supported by research which showed companies from the same business sectors tend to exhibit similar stock price fluctuations~\cite{gopikrishnan2000sector_price}.

\subsection{Generating Training Data}

\begin{figure}[t]
    \centering
    \includegraphics[width=0.85\columnwidth]{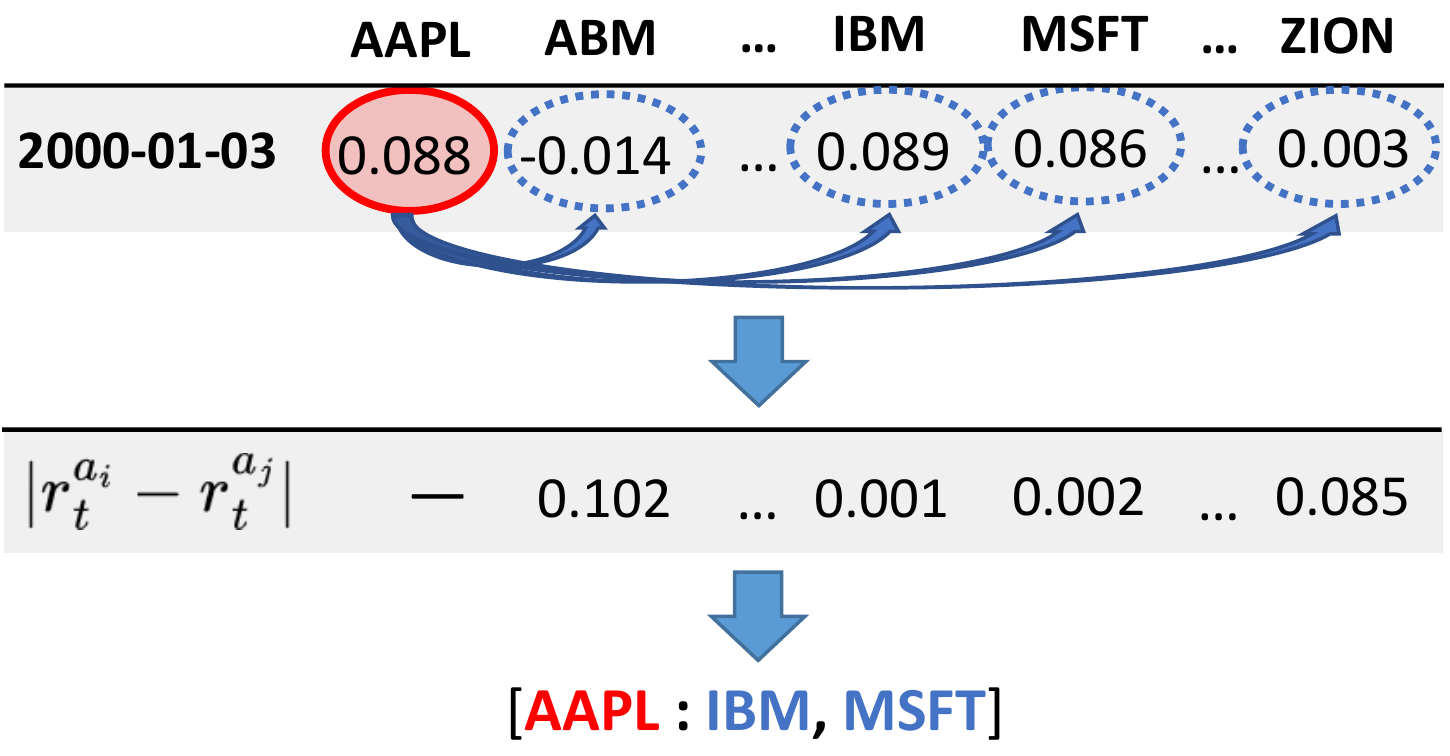}
    \caption{Generating training data, i.e., target:context stock sets}
    \label{fig:context}  
\end{figure}

Consider a universe of stocks $U = \{a_1,...,a_{|U|}\}$ and for each stock $a_i$ we have a vector $\mathbf{p}_{a_i}=\{p_0^{a_i},...,p_T^{a_i}\}$ containing its prices at discrete time intervals (daily or weekly for example). From the pricing data we then compute a returns vector $\mathbf{r}_{a_i}=\{r_1^{a_i},...,r_T^{a_i}\}$ using Equation \ref{eqn:returns_formula}.
\begin{equation}
    \label{eqn:returns_formula}
    r_{t}^{a_i} = \frac{p_{t}^{a_i}-p_{t-1}^{a_i}}{p_{t-1}^{a_i}}
\end{equation}

From these returns vectors we can generate \emph{target:context sets}. For a context size $C$, the context stocks for target asset $a_i$ at time $t$ are the $C$ stocks which have the closest return at that point in time. The closest return is defined by the lowest absolute value difference in return for candidate stock $a_j$, formulated as $|r_t^{a_i} - r_t^{a_j}|$. An example of this process is outlined in Figure \ref{fig:context}. Here we have AAPL as the context stock and $t$ is January 3\textsuperscript{rd} 2000. We compute the absolute value difference between the return of AAPL at that point in time with the return of each other stock at the same point in time. Then, we choose the $C$ stocks with the lowest values as the context stocks, excluding AAPL itself. In this case, IBM and MSFT have very small differences with AAPL and so are likely to be chosen as context stocks.
In total, we generate a target:context set for every stock at each point in time, which results in a total of $|U|\times T$ sets for training.

An example of a target:context set for $C=3$ might be $\mathcal{S}(a_{0},t)=[a_{0} : a_{270}, a_{359}, a_{410}]$, which corresponds to \emph{[AAPL : IBM, MSFT, ORCL]}, since we store the index value of the stock rather than strings (270 is the index corresponding to IBM in our universe of stocks, for example). This tells us that, at a certain point in time $t$, the three stocks with the closest returns to Apple Inc. were IBM, Microsoft and Oracle.%\todo[inline]{what about defining a general equation for target:context, and then use this definition in Figure 2 and Section Model Architecture, to replace x and y; For example, $[a_t : a_{c_1}, a_{c_2}, .., a_{c_C}]$}

% From these returns vectors we can generate \emph{target:context sets} for a given context size $C$. The target:context set $\mathcal{D}(a_i,t,C)$ for target stock $a_i$ at time $t$ contains the indices of the $C$ stocks that exhibit the closest return to the target stock at that point in time. For example, to generate $\mathcal{D}(a_i,t,C=3)$ we compute $|r_t^{a_i} - r_t^{a_j}|$ for all indices $j\neq i$ and store the three $j$ values that resulted in the smallest difference.

% \begin{equation}
%     \mathcal{D}(a_i,t,C) = \{j : |r_t^{a_i} - r_t^{a_j}| \text{ in } C \text{ smallest}\}
% \end{equation}

% For example, if we take $C=1$ we have $\mathcal{D}(a_i,t,C=1)$ simply equal to the set of size one containing the index of the stock which has the closest return to $a_i$ at time $t$. Mathematically, we define this as
% %-- Mathematical definition maybe not necessary here?
% %- If removed still must include |r_t^{a_i} - r_t^{a_i}| somewhere so that it is clear we are talking about closest = euclidean distance
% \begin{equation}
%     \mathcal{D}(a_i,t,C=1) = \argmin_{j\in J} |r_t^{a_i} - r_t^{a_j}|
% \end{equation}
% where we define $J=\{1,2,...,|V|\}\setminus\{i\}$ to be the set of all indices excluding that of the target stock.

\subsection{Base Model Architecture}\label{sec:mod_arch}

\begin{figure*}
    \centering
    \includegraphics[width=1\linewidth]{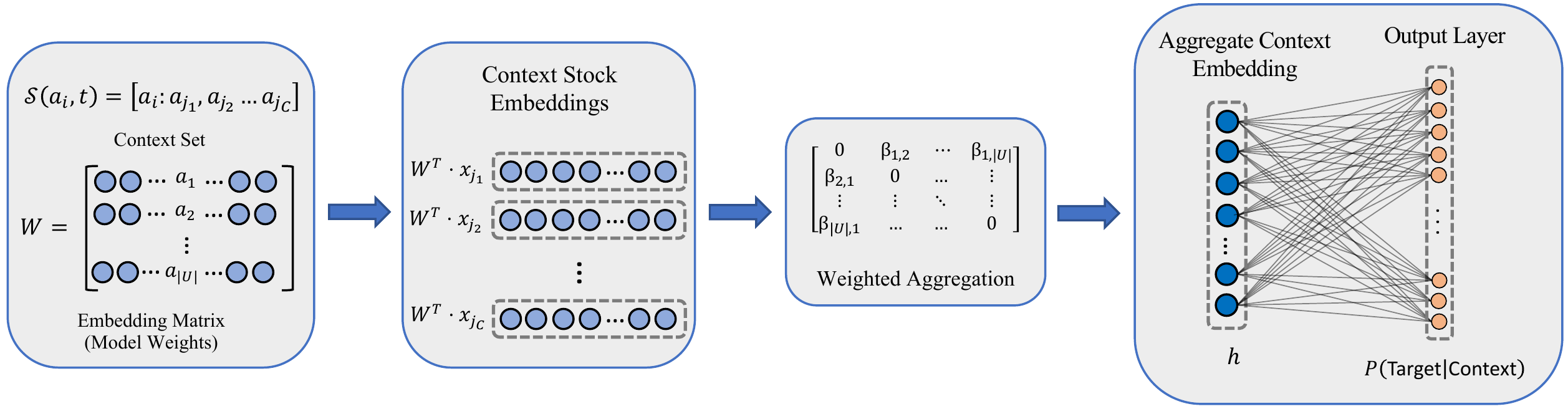}
    \caption{Model Architecture}
    \label{fig:CBOS}
\end{figure*}

% \begin{figure}
%     \centering
%     \includegraphics[width=0.7\columnwidth]{Images/architecture_upd.png}
%     \caption{Base Model Architecture}
%     \label{fig:CBOS}
% \end{figure}

The proposed model architecture is illustrated in Figure \ref{fig:CBOS}. The model design is such that \emph{the stock embeddings are the model parameters}. As such, each row in the weight matrix $\mathbf{W}\in \mathbb{R}^{|U|\times N}$ is a stock embedding, which are randomly initialised.

The first step is to compute the hidden layer, which is an element-wise average of the context stock embeddings. To be more precise, the input to the model is a one-hot encoded version of the context set, and so, consists of $C$ one-hot vectors $\{\mathbf{x}_{j_1},\mathbf{x}_{j_2}, ... , \mathbf{x}_{j_C}\}$, one for each context stock. These vectors are used to extract the embeddings corresponding to the $C$ context stocks. For example, computing $\mathbf{W}^T\cdot\mathbf{x}_{j_1}$ will extract a single row from $\mathbf{W}$ --- the embedding corresponding to the first context stock $a_{j_1}$. In its most basic form, the hidden layer, $\bm{h}$, is an element-wise average of the extracted embeddings, and is formulated in Equation \ref{eqn:hidden_layer}. %$\bm{h}$ has the same size as an embedding and can be thought of as a embedding representing the average of all the context stocks.

% where $\mathbf{x}_1\in \mathbb{R}^{|U|}$, for example, contains zeros everywhere except a 1 at the position corresponding to the index which had the most similar return to the target stock at that point in time. In other words, these input vectors are simply an expanded version of the context:target sets we defined previously.

% To compute the hidden layer $h$ we first compute $\mathbf{W}^T \cdot \mathbf{x}_j$ for each $j=1,2,...,C$. Since $\mathbf{x}_j$ is a one-hot vector this operation simply extracts a row from the matrix $\mathbf{W}$ which corresponds to a stock embedding. The final step to compute the hidden layer is simply to average across the $C$ stock embeddings. This series of operations is formulated in Equation \ref{eqn:hidden_layer}.

\begin{equation}
    \label{eqn:hidden_layer}
    \bm{h} = \frac{1}{C} \mathbf{W}^T (\mathbf{x}_{j_1}+\mathbf{x}_{j_2}+ ... + \mathbf{x}_{j_C})
\end{equation}

% \begin{equation}
%     \label{eqn:new_hidden_layer}
%     \bm{h} = \mathbf{W}^T (w_{i,j_1,t}\mathbf{x}_{j_1}+w_{i,j_2,t}\mathbf{x}_{j_2}+ ... + w_{i,j_
%     C,t}\mathbf{x}_{j_C})
% \end{equation}

Thus the hidden layer, $\bm{h}$, is an $N$-dimensional vector and can be thought of as an aggregate embedding representation of the context stocks. In Equation \ref{eqn:hidden_layer}, each embedding receives an equal weighting of $C^{-1}$; however, in Section \ref{sec:noise_reduction} we describe a more complex weighting approach intended to reduce noise. 

The next step is to estimate the probability of the target stock \emph{given} $\bm{h}$ by applying Equation \ref{eqn:CBOW_out}.

\begin{equation}
    \label{eqn:CBOW_out}
    \mathbb{P}(\text{Target }|\text{ Context}) = softmax(\mathbf{W} \bm{h})
\end{equation}

Ensured by using the softmax activation, the output is a posterior probability distribution expressing the probability of each stock in the universe being the target stock, given the context stocks observed. Since the dot product represents a measure of similarity between vectors, the model assigns higher probability to stocks whose embeddings are similar to hidden layer embedding $\bm{h}$. In this way, when back-propagation is applied, stocks which commonly co-occur in target:context sets will end up closer in the embedding space. As a result, the embeddings will learn nuanced relationships that are present in the historical returns data. We note that the ground truth here, $\mathbf{y}_i$ in Figure \ref{fig:CBOS}, is a one-hot vector of length $|U|$ indicating the true target stock.

\subsection{Noise Reduction Strategies}\label{sec:noise_reduction}

Financial returns data are notoriously noisy~\cite{de1990noise}, and so, in addition to the base model architecture, we propose two amelioration strategies to improve performance. Firstly, a weighting strategy based on overall distributional co-occurrence is introduced. With this, the hidden layer $\bm{h}$ is computed via a weighted average, and is implemented by scaling each $\mathbf{x}_j$ using a weight, which is proportional to the rate at which the given context stock $a_j$ appears in the context of the target stock $a_i$ over the whole training dataset. This is outlined in Equation \ref{eqn:weight_proportional}, where $\mathlarger{\mathds{1}}$ denotes the indicator function.

\begin{equation}
    \label{eqn:weight_proportional}
    w_{i,j,t} \propto \beta_{i,j} = \frac{1}{T}\sum_{t=1}^T \mathlarger{\mathds{1}}\Big( a_j\in \mathcal{S}(a_i,t)\Big)
\end{equation}
The constant of proportionality here, $k_{i,t}$, is computed such that the weightings over all context stocks sum to one. 
\begin{equation}
    \label{eqn:proportionality}
    w_{i,j,t}=k_{i,t}\cdot\beta_{i,j} \hspace{0.5cm}:\hspace{0.5cm} k_{i,t}=\frac{1}{ \sum_{j:j\in\mathcal{S}(a_i,t)}\beta_{i,j}}
\end{equation}

Secondly, the distribution of returns over a short time period, such as daily, contains a large proportion of values close to 0, which indicates little movement in stock price. In an effort to isolate meaningful cases and reduce noise, a context set $\mathcal{S}(a_i,t)$ was deleted from the training data if the target stock return, $r_t^{a_i}$, was within the interquartile range (IQR) of returns on that day. As a result, only sets where the target stock had a movement significantly different from the market average on a given day were included in training.

%==================================================
%==================================================
%==================================================
\section{Case Studies}\label{sec:case_studies}
In this section, for illustrative purposes, we describe two ways in which the distributed representations that can be learned may be used in financial settings to: (1) to identify nearest neighbour stocks; and (2) to complete useful stock analogies similar to those used to in the original Word2Vec work \cite{mikolov2013efficient}.
% Before presenting a two-pronged evaluation in Section \ref{sec:evaluation}, we present a number of case studies that showcase some interesting properties of the learned embeddings. %First, we show that stocks which lie close together in the embedding space tend to have similar business models. Secondly, we demonstrate how vector operations can be used to manipulate the embeddings and extract meaningful analogies. Finally, we outline 

% \footnote{A subset of the linked data was used, with the inclusion criterion that a stock had complete pricing data from 2000 onwards. \href{https://www.kaggle.com/ehallmar/daily-historical-stock-prices-1970-2018/}{https://www.kaggle.com/ehallmar/daily-historical-stock-prices-1970-2018/}}

\subsection{Daily Returns Dataset}
For these case studies, and the evaluation that follows in Section \ref{sec:evaluation}, we use a publicly available dataset of daily pricing data\footnote{All data and code is available at
%\href{https://github.com/anonymous}{https://github.com/anonymous}
\href{https://github.com/rian-dolphin/stock-embeddings}{https://github.com/rian-dolphin/stock-embeddings}
} for 611 US stocks during the 2000-2018 period. In addition to daily returns, each stock is also associated with a \emph{sector} and \emph{industry} classification label. The former corresponds to the business sector the company operates in and there are eleven sectors in total, including Finance, Health Care and Technology, for example. The industry label represents a finer-grained classification; a stock in the Technology sector may have Computer Software as its industry label, for example, to contrast it with another Technology stock in the Electronic Components industry.

%It is worth noting that although our experiments used data only from US listed equities, this technique is applicable to any group of financial assets whose pricing information is available. As a result, we can apply the proposed methodology to a wider universe involving multiple asset classes such as fixed income, options and other derivatives.

%==================================================
%==================================================
\subsection{Identifying Nearest Neighbour Assets}

\begin{table*}[]
    \centering
    \caption{Examples of Top-3 Nearest Neighbours for Given Query Stocks}
    \label{tab:KNN}
    \begin{tabular}{cccc}
        \textbf{\begin{tabular}[c]{@{}c@{}}Query Stock\\ Sector - Industry\end{tabular}} & \textbf{3 Nearest Neighbours - Sector - Industry} & \textbf{Similarity}  \\ \midrule
        \begin{tabular}[c]{@{}c@{}}JP Morgan Chase\\ Finance \\Major Bank\end{tabular}          & \begin{tabular}[c]{@{}c@{}}Bank of America Corp - Finance - Major Bank \\ State Street Corp - Finance - Major Bank  \\ Wells Fargo \& Company - Finance - Major Bank\end{tabular}  & \begin{tabular}[c]{@{}c@{}c@{}}0.88\\0.82\\0.81\end{tabular} \\\hline
        
        \begin{tabular}[c]{@{}c@{}}Analog Devices\\ Technology \\ Semiconductors\end{tabular}          & \begin{tabular}[c]{@{}c@{}}Maxim Integrated - Technology - Semiconductors \\ Texas Instruments - Technology - Semiconductors  \\ Xilinx, Inc. - Technology - Semiconductors \end{tabular} & \begin{tabular}[c]{@{}c@{}c@{}} 0.93 \\ 0.91 \\ 0.90\end{tabular} \\ \hline
        
        \begin{tabular}[c]{@{}c@{}}Chevron Corporation\\ Energy \\ Integrated Oil \& Gas\end{tabular}          & \begin{tabular}[c]{@{}c@{}}Exxon Mobil - Energy - Integrated Oil \& Gas \\ BP P.L.C. -  Energy - Integrated Oil \& Gas \\ Occidental Petroleum - Energy - Oil \& Gas Production \end{tabular} & \begin{tabular}[c]{@{}c@{}c@{}}0.89\\0.82\\0.78\end{tabular}  \\ \hline
    \end{tabular}
\end{table*}

Ideally, we should expect the distributed representations of related stocks to to be closer, by some suitable similarity metric, than dissimilar stocks, and the ability to identify similar (and dissimilar) stocks is an important tool for portfolio design. Here we use cosine similarity as our similarity metric in order to provide examples of the \emph{k} nearest neighbours for a sample of example stocks; similar results can be obtained when using alternative metrics such as Euclidean distance.

% The first case study involves a nearest neighbours analysis, a natural first point of reference to sanity check any latent space representations. We would hope that stocks which appear very close together in the latent space should be related somehow. In order to find the $k$-nearest neighbours (kNN) for a given query stock we first must define what exactly we mean by nearest. Here we implement kNN using Euclidean distance as the similarity metric. It is worth noting that, unsurprisingly, the same pattern of related stocks emerges if we use cosine similarity or dot product. We would expect this as the model architecture uses the dot product as the internal measure of similarity.

Table \ref{tab:KNN} shows the top-3 ($k=3$) nearest neighbours for JP Morgan Chase, Analog Devices, and Exxon Mobil Corp, three well-known companies in very different sectors. In each case the nearest neighbours pass the \say{sanity test} in that they belong to similar sectors and industries. For example, the three nearest neighbours of JP Morgan, a major bank, are also all major banks. Remember, that no sectoral or industry information has been used in determining these nearest neighbours, because only daily returns information has been used to generate the distributed representations used for similarity assessment.

% a number of query stocks. For example, taking JP Morgan Chase, a well-known financial institution that is classified in the Finance sector and Major Banks industry, we find all three of its nearest neighbours are also well-known financial institutions classified in the same sector and industry. 
%\begin{tabular}[c]{@{}c@{}c@{}}0.88//0.82//0.81\end{tabular}
%\todo[inline]{put the sector information of Nearest Neighbours into the table}

% This interesting result is echoed throughout the dataset. It is worth highlighting again that the data used to train the embeddings is comprised purely of historical returns data. As a result, these observed examples provide experimental evidence to support our initial hypothesis --- related stocks disproportionately exhibit similar returns at the same time.

Nearest neighbour stocks, can be used in a variety of ways by investors. By focusing on the $k$ nearest neighbour stocks we can develop a basic stock recommendation system which, when given a target stock -- a novel stock for the investor or one already in their portfolio -- can generate a ranked list of similar stocks based on their historical returns data. Conversely, the ability to identify maximally dissimilar stock is an important way to improve portfolio diversity in order to provide an investor with an ability to guard against volatility and \emph{hedge} against sudden sectoral shocks.

% which tend to have the same returns patterns is of great interest to investors. One could use the embeddings as the basis for a stock recommendation system whereby an investor could input stocks they are interested in investing in and receive some similar companies with which to carry out a comparable company analysis. Additionally, an investor may be interested to input a stock already in their portfolio. By doing so they can identify similar stocks which would allow them to gain exposure to the business sector they believe to be promising while reducing company specific (idiosyncratic) risk, a concept known as diversification.

%==================================================
%==================================================
\subsection{Analogical Inference in Financial Domains}

In the NLP domain, one of the most appealing and useful features of word embeddings has been their ability to capture relationships that go deeper that simple syntactic regularities. Word embeddings have provided access to a form of analogical inference using simple vector operations over the embeddings \cite{ethayarajh2018towards}. The classic example is the analogy \say{Man is to King as Woman is to ?} which can be solved algebraically as \textit{vector}(\say{King}) - \textit{vector}(\say{Man}) + \textit{vector}(\say{Woman}) to produce a vector that is closest to the vector representation of the word Queen~\cite{mikolov2013efficient}. 

Similar algebraic operations can be used in the financial domain, albeit in a manner that is not quite as intuitive as the word analogies that we are all so familiar with from everday life. For example, in Table \ref{tab:analogies} we solve for similar analogies to identify, for example, Hewlett Packard is the stock in the Technology sector that is most similar to American Express in the Finance sector, or that Johnson \& Johnson in Healthcase is most similar to Exxon Mobil in Energy. Such analogies can be very useful in an investment context when investors wish to extend their portfolios into different sectors in which they have little or no investment experience and/or to improve their portfolio diversity and limit overall risk and volatility.

\begin{table}[]
    \centering
    \caption{Analogical Examples}
    \label{tab:analogies}
    \begin{tabular}{p{5cm} c}
    \multicolumn{1}{c}{\textbf{Analogy}}                               & \multicolumn{1}{c}{\textbf{Similarity}} \\ \hline
    American Express is to Finance as Hewlett Packard is to Technology & 0.8                                     \\ \hline
    Exxon Mobil is to Energy as Johnson \& Johnson is to Health Care      & 0.81                                    \\ \hline
    State Street is to Finance as Xerox is to Technology               & 0.8   
            \\ \hline                                 
    \end{tabular}
    
\end{table}

% . Consider, for example, a stock recommendation application where a trading platform wishes to recommend investments to its customers. A service could take existing investments that the customer has chosen and use these to find companies in other sectors that might suit a customer's investment philosophy, risk appetite etc. These companies would offer attractive investments with which the customer could increase their diversification, thus reducing their idiosyncratic risk.

%==================================================
%==================================================
%==================================================

%\subsection{Differences from Word2Vec}

%Do we need this section?

%Since the approach is similar to W2V it might be worth mentioning the differences?

%==================================================
%==================================================
%==================================================
\section{Evaluation}\label{sec:evaluation}
In this section we provide the results of an initial evaluation of the proposed stock embeddings approach. We describe two experiments to evaluate different aspects of the learned distributed representations. In the first, we evaluate the ability of the distributed representations to be used for sector classification. In the second, we describe and evaluate an approach to generate hedged portfolios that benefit from lower volatility characteristics than more conventional hedging strategies, which rely on simple price correlations. In both experiments we use the daily returns dataset described previously and generate our embeddings with a context size of 3 and an embedding dimension of 20. Both noise reduction techniques (distributional co-occurrence and significant movement) were also used in training.

% To further present the efffenciency of learning stock embeddings we will describe a number of experiments in two classical fianacial tasks and show interesting case studies that convey the useful information captured by our embeddings. For the entirety of this section the embeddings were trained as described in Section \ref{sec:Methodology} with a context size of 3 and embedding dimension of 20. The data used is the same as that described in Section \ref{sec:case_studies}.

%==================================================
%==================================================
%==================================================

\subsection{Sector Classification}
The complex web of hidden factors and unpredictable events that influence stock prices make investing an undeniably risky endeavour. In particular, investing in individual stocks exposes the investor to both broader market risk (systematic) and asset-specific risk (idiosyncratic). Exchange traded funds (ETFs) have become a popular alternative to stock picking since the \emph{diversification} they offer greatly reduces idiosyncratic risk and their cost is relatively low. 

ETFs are securities traded on public exchanges that offer partial ownership in a portfolio containing a large number of stocks. Often these portfolios are constructed in such a way that they aim to track a particular geographical region or specific market sector so that an investor can obtain highly diversified exposure to any market segment that they wish. ETFs have become so popular that, by the end of 2016, their market share exceeded 10\% of the total market capitalisation traded on US exchanges, representing more than 30\% of overall trading volume~\cite{ben2016exchange}.

However, companies offering ETF products need to decide on the constituents of the underlying portfolios. Sometimes the portfolios are constructed to track certain indices, such as the S\&P500. However the selection criterion can be highly subjective, particularly for market sector ETFs. For example, a company like Amazon is classified as \say{consumer discretionary} by the Global Industry Classification Standard (GICS). Should it be included in a consumer discretionary ETF with the likes of Ford Motor and McDonald's? Or would it be better suited to a technology or consumer staples ETF? Indeed, a case can be made for Amazon to be considered as all of the above.

The ability to segment stocks into market sectors is also important for many other types of financial and economic analysis --- measuring economic activity, identifying peers and competitors, quantifying market share and bench-marking company performance --- none of which would be possible without industry classifications~\cite{phillips2016industry}. The merit of particular assets are primarily determined relative to their peers within the same asset class and business sector. Thus, a well-defined sector classification system makes it easier for analysts to compare companies’ relative valuations and to build sector-specific return and risk estimates~\cite{GICS2020}.

With these use cases in mind, we will show how the stock embeddings can be used to successfully classify stocks into business sectors and pick up on inconsistencies in existing classification schemes. First, we explore a visualisation of the latent space spanned by the learned embeddings, examining clustering and relationships between stocks. Then we consider some interesting cases which seem, on the surface, not to behave how we might expect. Finally, we train a classification model based on the embeddings which assigns sector labels to companies.

%==================================================
%==================================================
%==================================================

\subsubsection{Clustering Stocks using Embeddings}
%As previously mentioned, our hypothesis that \emph{financial assets that disproportionately exhibit similar returns at the same time are related} is supported by existing research\cite{gopikrishnan2000sector_price}; however, in this section we provide further evidence by showing that

% From the learned embeddings, each of which is a $N=20$ dimensional vector, we can obtain lower dimensional representations using dimensionality reduction techniques such as principal components analysis (PCA). These PCA representations can then be visualised and checked for initial clustering and patterns.

% \begin{figure}[]
%   \centering
%   \includegraphics[width=\linewidth]{Images/PCA_3D.png}
%   \caption{A 3-Dimensional Visualisaion of Stock Embeddings Coloured by Business Sector, with Edges Indicating Cosine Similarity $>0.8$}
%   \label{fig:3D_PCA}
% \end{figure}

% Figure \ref{fig:3D_PCA} shows a 3D visualisation of the embeddings for stocks in three of the largest business sectors: Technology, Finance and Public Utilities. Each node represents a stock, colored by sector, and an edge indicates that the two nodes it connects have greater than 0.8 cosine similarity between their embeddings. The clustering of stocks into business sectors is clearly evident in the plot. We can also see that nearly all of the edges in the graph are between nodes from the same sector. From this we conclude that the model architecture and training procedure result in embeddings that have successfully picked up on relationships between stocks. 

\begin{figure}[]
  \centering
  \includegraphics[width=\linewidth]{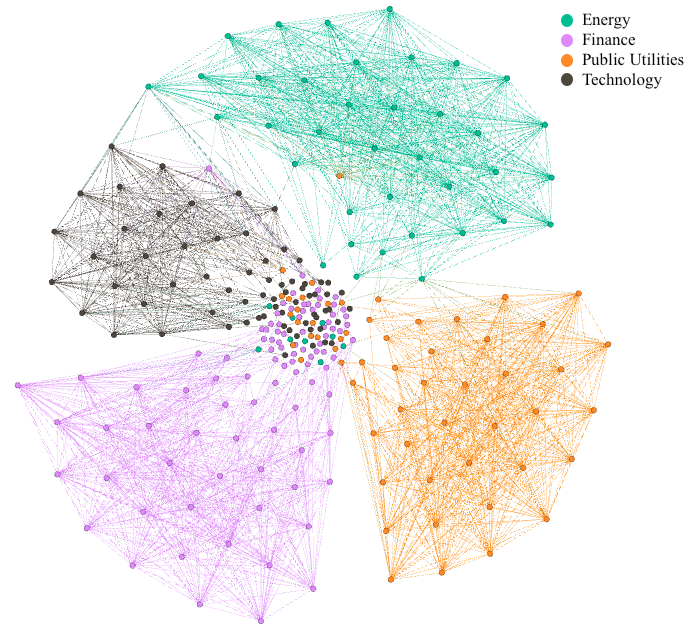}
  \caption{Graphical Visualisaion of Stock Embeddings Coloured by Business Sector. Edges Formed by an Embedding Similarity Threshold}
  \label{fig:gephi}
\end{figure}

Visualising latent embeddings in a lower dimensional space can often be useful to identify relationships and clustering behaviour. Figure \ref{fig:gephi} shows a graphical representation of the embeddings for stocks in four of the largest business sectors: Energy, Finance, Public Utilities and Technology. Each node represents a stock, colored by sector, and an edge indicates that the two nodes it connects have greater than 0.7 cosine similarity between their embeddings. The plot is generated using a force-directed graph drawing algorithm in Gephi\cite{ICWSM2009gephi}.

The clustering of stocks into business sectors is clearly evident in the plot. We can also see that nearly all of the edges in the graph are between nodes from the same sector. From this we conclude that the proposed model architecture and training procedure result in embeddings that have successfully picked up on relationships between stocks.

When we consider that the training data used is purely derived from historical returns data, this is a very positive result because it suggests that it is possible to reconstruct important sectoral information from the embeddings, and indeed are likley to be able to do so in a way that is more nuanced than might be possible using simple sectoral labels.

% \begin{figure}[htb!]
%   \centering
%   \includegraphics[width=\linewidth]{Images/PCA_2D.png}
%   \caption{A 2-Dimensional Visualisaion of Stock Embeddings Coloured by Business Sector}
%   \Description{2-Dimensional Visualisaion of Stock Embeddings Coloured by Business Sector. Clusters are seen to form.}
%   \label{fig:2D_PCA}
% \end{figure}

%Figure \ref{fig:2D_PCA} shows a two-dimensional projection of the embeddings obtained using principal components analysis (PCA) for dimensionality reduction. Four of the largest business sectors are shown; technology, finance, public utilities and energy. We see clear clustering of stocks by sector. This is a very interesting observation when we consider that our embeddings are trained using sets stemming from only historical returns data and nothing else. 

%In Figure \ref{fig:3D_PCA} we again see a lower dimensional representation of stocks computed using PCA applied to their embeddings. Each node on the graph represents a stock and edges between nodes indicate a cosine similarity between embeddings greater than 0.8. Three sectors are included here and, in addition to the clustering which is evident again, we can see lots of edges between stocks from the same business sector. Though there are some stock nodes which reside somewhat between clusters, these nodes tend not to have any edges connecting to other stocks.

% \begin{figure*}
%     \includegraphics[width=.45\textwidth]{Images/PCA_2D.png}\hfill
%     \includegraphics[width=.45\textwidth]{Images/PCA_3D.png}
% \end{figure*}

%==================================================
\subsubsection{High Similarity Mismatches}\label{sec:mismatch}

\begin{table*}[]
    \centering
    \caption{Examples of high similarity mismatches --- stocks with very high similarity that have different sector labels}
    \label{tab:mismatch}
    \begin{tabular}{ccc}
        \textbf{\begin{tabular}[c]{@{}c@{}}Stock A \\ Sector - Industry\end{tabular}}                         & \textbf{\begin{tabular}[c]{@{}c@{}}Stock B\\ Sector - Industry\end{tabular}}                     & \textbf{Similarity} \\ \midrule
        \begin{tabular}[c]{@{}c@{}}Lennar Corporation\\ Basic Industries - Homebuilding\end{tabular}          & \begin{tabular}[c]{@{}c@{}}KB Home\\ Capital Goods - Homebuilding\end{tabular}                   & 0.98                \\ \hline
        \begin{tabular}[c]{@{}c@{}}Bristow Group Inc.\\ Transportation - Transportation Services\end{tabular} & \begin{tabular}[c]{@{}c@{}}Unit Corporation\\ Energy - Oil \& Gas Production\end{tabular}        & 0.92                \\ \hline
        \begin{tabular}[c]{@{}c@{}}Cirrus Logic, Inc.\\ Technology - Semiconductors\end{tabular}              & \begin{tabular}[c]{@{}c@{}}Xcerra Corporation\\ Capital Goods - Electrical Products\end{tabular} & 0.93                \\ \hline
    \end{tabular}
\end{table*}

Though nearly every edge in the Figure \ref{fig:gephi} occurs between nodes from the same sector, there are some cases where this is not true. In other words, there are some cases where two stocks have a very high embedding similarity but their sector labels don't match. Does this highlight a flaw in the embeddings, where stocks achieve very high embedding similarity when they should not? To answer this we provide some examples of these \emph{high similarity mismatches}. We consider pairs of stocks that have a cosine similarity between embeddings greater than 0.9 and are members of different sector classes. A number of examples are shown in Table \ref{tab:mismatch}. 

The first example in Table \ref{tab:mismatch} is Lennar Corporation. Lennar is a home construction and real estate company which has been recorded as a member of the Basic Industries sector and the Homebuilding industry (remember here that industry is a finer-grained classification of the sector). Lennar Corporation actually has four stocks that have greater than 0.9 similarity but are \emph{not} in the Basic Industries sector, we can see one of these in Table \ref{tab:mismatch}. KB Home is classified in the Capital Goods sector but we note that it is also in the Homebuilding industry. In fact, these companies both operate a very similar business model focused around home construction. This is the same for the other three stocks which have very high similarity with Lennar; all are in the Homebuilding industry but were subjectively classified into a different sector to Lennar.

Immediately we can see that the human approach of assigning sector classifications can be flawed and inconsistent. Why is a company in the same industry but not the same sector when industry is just a finer grained description of sector? As previously mentioned, sector classification is a vital task in the financial domain for many reasons. We propose that the embeddings could be a useful and objective tool to complement the existing methods by which companies are grouped.

Another example is the case of Cirrus Logic and Xcerra Corporation. Though these companies are segmented into a different sector and industry, both of their business models revolve around dealing with semiconductors, Xcerra develops equipment for testing semiconductors and Cirrus is a semiconductor supplier. Similarly, Unit Corporations engage in oil and gas production while Bristow Group offers offshore oil and gas transportation.

These two examples are not labelling inconsistencies in the same way as the Homebuilding example, but rather, they indicate that the embeddings pick up on interesting relationships which are not captured in the sector classification schemes. This particular property of the embeddings, to pick up on nuanced relationships between stocks, will be explored further in Section \ref{sec:hedging}.

%This further supports that an automated industry classification/index creation tool would be very useful in practice. Additionally, it indicates that the sector classification task we attempted in the previous subsection is even more difficult that it may first seem, since companies which are actually operating in the same business industry may be subjectively classified into different sectors. An analysis like this, looking at high similarity mismatches, could be very useful for companies like Standard \& Poors who maintain industry classification schemes, to identify inconsistencies in their classification.

%There are other examples like Lennar Corporation. In fact, every case of very high similarity but differing sector plays out like the example above, where the two companies are in fact running very similar operations but just happen to have been classified into different sectors.

%==================================================
%==================================================
%==================================================
\subsubsection{Sectoral Classification}

Using the embeddings generated by the proposed model architecture we can use a classification model to segment companies into business sectors in an objective manner, based purely on relationships in historical returns. To do this we train a classification model with embeddings as input and sector label as the output.

There are a number of considerations here which will undoubtedly limit the accuracy of the classification model. Firstly, the training data is derived solely from historical returns data, which are influenced by a complex network of unpredictable factors. Secondly, as we saw in Section \ref{sec:mismatch} when looking at high similarity mismatches, there are a number of stocks which are either labelled inconsistently or subjectively, and could be classified into a number of different sectors. As a result, sector classification in the financial domain is a very challenging problem and even more so when using only returns data.

To train the classification model we used $k$-fold cross validation with $k=5$. Within the data there is a large class imbalance between sectors which can introduce algorithmic bias and impact negatively on results~\cite{blanzeisky2021algorithmic}. And so, to remedy this we applied Synthetic Minority Oversampling Technique (SMOTE) to the training data. As a classification model we use a support vector classifier.%~\cite{platt1999SVM}. 

\begin{table}[]
    \centering
    \caption{Results from the sector classification model which takes embeddings as inputs}\label{tab:classification_results}
    \begin{tabular}{ccccc}
    \toprule
        \textbf{Model}                               & \textbf{Precision} & \textbf{Recall} & \textbf{F1} & \textbf{Accuracy} \\
        \midrule
        Embedding                                    & 0.57               & 0.54            & 0.55        & 54\%              \\
        Embedding + IQR                              & 0.59               & 0.56            & 0.56        & 56\%              \\
        Embedding + Weight                           & 0.59               & 0.57            & 0.57        & 57\%              \\
        \multicolumn{1}{l}{Embedding + Weight + IQR} & \textbf{0.62}               & \textbf{0.60}            & \textbf{0.60}        & \textbf{60\%}      \\
        \bottomrule
    \end{tabular}
\end{table}

% \begin{table}[]
%     \centering
%     \caption{Results from the sector classification model which takes embeddings as inputs}
%     \label{tab:classification_results}
%     \begin{tabular}{lcccc}
%     {Sector} &  \textbf{Precision} &  \textbf{Recall} &  \textbf{F1} &  \textbf{Support} \\
%     \midrule
%     Basic Industries      &       0.43 &    0.77 &      0.56 &       13 \\
%     Capital Goods         &       0.64 &    0.39 &      0.49 &       23 \\
%     Consumer Durables     &       0.56 &    0.42 &      0.48 &       12 \\
%     \begin{tabular}[l]{@{}l@{}}Consumer\\ Non-Durables\end{tabular} &       0.40 &    0.40 &      0.40 &       15 \\
%     Consumer Services     &       0.68 &    0.63 &      0.65 &       27 \\
%     Energy                &       0.88 &    0.93 &      0.90 &       15 \\
%     Finance               &       1.00 &    1.00 &      1.00 &       30 \\
%     Health Care           &       0.67 &    0.89 &      0.76 &        9 \\
%     Public Utilities      &       0.94 &    0.79 &      0.86 &       19 \\
%     Technology            &       0.73 &    0.84 &      0.78 &       19 \\
%     Transportation        &       0.50 &    0.50 &      0.50 &        2 \\
%     \midrule
%     Weighted Avg          &       0.72 &    0.71 &      0.71 &      184 \\
%     \midrule
%     Overall Accuracy              &       71\% &     &       &         \\
    
%     \bottomrule
%     \end{tabular}
% \end{table}

Table \ref{tab:classification_results} shows the performance in the sector classification task. The highest accuracy of 60\% is achieved by the Embedding model with both noise reduction techniques, thus indicating their utility. This is an impressive result considering there are 11 sector classes\footnote{The sector classes are: Basic Industries, Capital Goods, Consumer Durables, Consumer Non-Durables, Consumer Services, Energy, Finance, Health Care, Public Utilities, Technology, Transportation} as well as the aforementioned limitations. A more in depth analysis shows a variation in accuracy across sectors, with the more populated sectors being very accurately classified (F1 $>0.9$ in some cases) while other minority sectors have quite low accuracy. With this in mind, we believe the accuracy could be improved with a larger dataset. 

An objective method of sector classification as described in this work has the potential to be very useful in practice to pick up on and prevent inconsistent company segmentation, which is a well-documented issue~\cite{chan2007industry}.

%[Address the fact there is no baseline here? I'm not entirely sure on the best way to discuss results here given nothing to compare to.]

%(Though no actual baselines exist here, another recent paper attempted to perform sector classifications using a model which utilised both returns data and textual data from a companies corporate filings. Their results were in line with this ~60\% despite using extra sources of data and large pre-trained language models. \cite{ito2020embedding})

%==================================================
%==================================================
%==================================================
\subsection{Hedging/Diversification Example}\label{sec:hedging}

As mentioned previously, much of the existing literature applying computational methods to the stock market has focused on returns forecasting. However, maximising returns is not the primary consideration for all investors; in fact, in some cases portfolio protection is a higher priority. For example, the primary focus of a portfolio manager for a defined benefit pension fund is having enough money to cover the agreed benefits. As a result, maximising returns in this context is a secondary consideration to successful risk management.
In order to achieve protection, investors and portfolio managers implement the ideas of diversification and hedging, and measure their effectiveness in terms of volatility. Thus it is not only important to be able to identify a set of similar stocks, but also a set of dissimilar stocks, which can be expected to behave in opposition to the similar stocks~\cite{dolphin2021measuring}. 
This is because, in a trading context, traders will often need to offset or hedge their positions in target stocks by also trading in maximally dissimilar stocks; the idea being that under-performance in a selected stock can be offset by gains in a dissimilar stock, thereby allowing a trader to limit their overall risk.

The usual practice is to implement hedging by using negatively-correlated assets leading to hedging strategies that rely on different types of correlation metrics. Here we consider an alternative approach by using the embeddings generated to identify maximally dissimilar stocks and inform a hedging strategy. For evaluation purposes we consider the case where an investor holds an open position in a given stock (the query stock) and wants to use another single stock (the hedge stock) to hedge and reduce their risk, measured as volatility, as much as possible.

\begin{algorithm}
\caption{Algorithm for finding hedged portfolios and simulating realised volatility}\label{alg:portfolios}
    \For{Query\_stock \textbf{in} stocks}{
    temp\_min $\gets \infty$\\
    Hedge\_stock $\gets$ None\\
        \For{temp\_stock \textbf{in} stocks}{
            \If{similarity(Query\_stock, temp\_stock) $<$ temp\_min}{
                temp\_min $\gets$ similarity(Query\_stock, temp\_stock)\\
                Hedge\_stock $\gets$ temp\_stock
            }
        
        }
    Simulate portfolio performance with equal weight between Query\_stock and Hedge\_stock and return the volatility. %(Hedge\_stock is the lowest similarity pair to Query\_stock found from algorithm above)\\
    }
\end{algorithm}

To implement this, we create a hedged two-asset portfolio for each stock in the dataset, as outlined in Algorithm \ref{alg:portfolios}. This results in a sample size of 611 portfolios, each of which contains two assets, the query stock and the hedge stock. The hedge stock is chosen as the lowest similarity counterpart to the query stock, where similarity is either cosine similarity between embeddings or one of the baselines which are listed later. We can then simulate the performance of the portfolios using each similarity metric in turn, recording the realised volatilities for comparison. %to achieve maximum risk reduction and thus minimise volatility, which is the goal here. 

\begin{figure}[]
  \centering
  \includegraphics[width=0.9\linewidth]{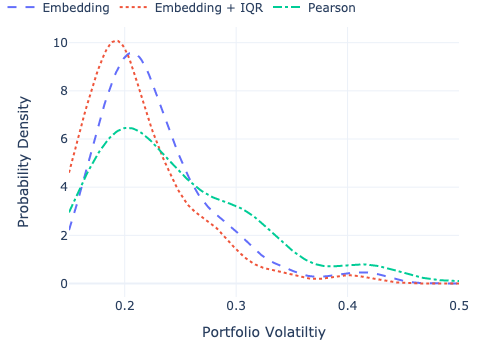}
  \caption{Distribution of Volatility for Hedged Portfolios}
  \label{fig:vol_density}
\end{figure}

\begin{table}[]
    \centering
    \caption{Portfolio hedging experiment results along with Tukey HSD test indicating significantly lower volatility than Pearson baseline at $\alpha=0.01$.}\label{tab:agg_hedge_results}
    \begin{tabular}{l|cc}
        \toprule
        \textbf{Method} & \textbf{Avg Volatility} & \textbf{Significant} \\
        \midrule
        Pearson                             & 23.8\%                  & -                    \\
        Spearman                            & 24.0\%                  & \xmark                    \\
        Geometric                           & 23.9\%                  & \xmark                    \\
        \midrule
        Embedding                           & 22.9\%                  & \cmark                    \\
        Embedding + IQR                     & \textbf{21.3\%}                  & \cmark                    \\
        Embedding + Weight                  & 22.8\%                  & \cmark                    \\
        Embedding + Weight + IQR            & 21.9\%                  & \cmark                    \\
        \bottomrule
    \end{tabular}
\end{table}

Of course, a train-test split is vital here to ensure that the experiment mimics a real-world out-of-sample trading application as much as possible. Given the time sensitive nature of financial data the split used a date cutoff, with the first 70\% of data (approx 2000 - 2013) being used to train the embeddings and compute the other baseline similarity metrics. The latter 30\% of data (approx 2013 - 2018) is then used to simulate the portfolios and compute realised volatility.

In modern portfolio theory~\cite{markowitz1952portfolio}, similarity is defined in terms of Pearson correlation and it is still the go-to method in both industry and academia. We propose that cosine similarity between the proposed embeddings may be a better choice. In the evaluation we include Pearson correlation as a baseline along with Spearman rank-order correlation coefficient and a recently proposed geometric shape similarity~\cite{chun2020geometric}.
%-- SAY HERE THAT THERE ARE NO ML BASELINES?
%- Not sure how to phrase

%Diversification is a simple idea but forms the foundation for modern portfolio theory. By holding multiple assets rather than just one, the risk of a portfolio is spread out and the investor is not completely exposed to the downturn in a single company. However, not all diversification is equal, we can attempt to selectively choose assets to obtain optimal diversification. In modern portfolio theory this selection is based on the Pearson correlation between asset returns; to achieve maximum diversification one should choosing assets with the lowest correlation of returns. There are, however, a number of limitations to this solution. Firstly, the theory is based on the underlying assumption that returns follow a normal distribution~\cite{markowitz1952portfolio}, but this has been shown not to hold true~\cite{}. Secondly, Pearson correlation has been shown to be a flawed similarity measure in the financial domain that is often misinterpreted and misunderstood~\cite{dolphin2021measuring, lhabitant2020correlationEDHEC}.

Figure \ref{fig:vol_density} shows the distribution of volatility over the 611 portfolios for the Pearson baseline, the proposed embeddings with no noise reduction and the best performing variation of the proposed architecture\footnote{All variations and baselines not included in Figure \ref{fig:vol_density} for visual clarity. See Table \ref{tab:agg_hedge_results} for more detail.}. From Figure \ref{fig:vol_density} we can see that the embedding approaches result in a greater concentration of lower volatility portfolios than the Pearson baseline, which has a fatter long-tail indicating a greater tendency towards high volatility portfolios. 
% have more probability density concentrated in the lower volatility regions, implying that these portfolios were better hedged. We also see that the embedding volatility has the least prominent right tail which indicates that the embedding similarity method generated fewer high volatility portfolios relative to the baselines.

Table \ref{tab:agg_hedge_results} displays the average volatility results. To ensure the robustness of results, we reran the experiment 100 times where, instead of choosing the single most dissimilar stock as the hedge stock, we randomly chose one of the 25 most dissimilar for each target stock on each iteration. Overall, the proposed embedding approach with IQR noise reduction results in portfolios with the lowest average volatility at 21.3\%. Post-hoc Tukey HSD tests indicate that the volatility in all of the embedding based methods is statistically significantly lower than the Pearson baseline at $\alpha=0.01$; none of the other approaches generate any statistically significant differences in mean volatility over the Pearson approach.

% In addition to a training and test split, we wanted to ensure that a small number of \say{luckily} chosen outliers did not skew the results. And so, instead of running the experiment only once and choosing the single least similar asset as the hedging stock each time, we repeated the experiment 100 times with the adjustment that when choosing the hedging stock for each asset we chose uniformly from the 25 least similar assets, rather than just the single least similar. In this way, we increased the robustness of the experiment and allowed for a significance test to be carried out.

Thus, we have shown how the proposed embeddings might be used to inform a hedging strategy that is superior to a number of baselines, at least within the simplified setting used for this experiment. Obviously real-world settings are based on more complex portfolios with many different stocks that need to be collectively hedged, and it is a matter for future work to further evaluate our embeddings approach in these more realistic settings. That being said, the approach used here still serves as an important indication of success: had the embeddings approach not be able to demonstrate improved volatility in these simple two-stock portfolios then it would cast doubts on its likely future success in more complex portfolios. Similarly, there is more work to be carried out when it comes to understanding the dynamics of the distributed representations and how they are learned: how does changing the context size or embedding dimension impact these findings, for example.

% This experiment is somewhat simplistic in that it explores the two-asset case (i.e. each portfolio only contains the query and hedge stock), but this is purposely so. Due to the small number of assets involved in each portfolio we can attribute our results to the selection method with higher confidence. With such positive results in the two-asset setting we believe that our proposed embedding similarity can be used to successfully inform hedging and diversification strategies in more complex $n$-asset portfolios.

%==================================================

\section{Conclusion and Future Work}

This work has focused on the problem of measuring relationships between the returns of financial assets. We proposed an approach to learning distributed representations for financial assets, based on daily returns data.%, that was inspired by recent work in language modeling \cite{mikolov2013efficient}. 
We demonstrated the utility of these representations using a number of example case-studies and by evaluating their efficacy in two important sub-tasks, sector classification and portfolio hedging. The results speak to the potential benefits of our approach and provide a useful starting point for further exploration and development.

In the future we plan to further evaluate the approach using more complex portfolio management settings and additional datasets. For example, the proposed technique is applicable to any group of financial assets whose pricing information is available, and so, we can expand the universe to include multiple asset classes in addition to equities, such as bonds, options and other derivatives. Additionally, we plan to carefully explore the parameter space of the model used to learn the embeddings, along with different approaches to generate context stocks. For example, if Apple's stock price rises on a certain day, the stock prices of its suppliers will often be affected in the days that follow. Can such \emph{lagged embeddings} be learned? Are they useful? And if so, how can they be used in practice?

\bibliographystyle{IEEEtran}
\bibliography{bibliography}

\end{document}